\documentclass{desyproc}

\begin{document}
\title{Completion of Phase I and Preparation for Phase II of the HAYSTAC Experiment}

\author{{\slshape Nicholas M. Rapidis$^1$}\\[1ex]
$^1$HAYSTAC Collaboration\\
University of California, Berkeley, Berkeley, USA}
 

\confID{20012}  
\desyproc{DESY-PROC-2018-03}
\acronym{Patras 2018} 
\doi  

\maketitle

\begin{abstract}
The HAYSTAC experiment utilizes a tunable resonant microwave cavity to search for dark matter axions. We report on the system and the results from Phase I of the experiment. This phase relied on a 9 T magnet, Josephson parametric amplifiers, and a dilution refrigerator for the operation of the experiment. Axion models with two photon coupling $\ g_{a\gamma\gamma}\gtrsim2\times 10^{-14}\ \mathrm{GeV}\ $ were excluded in the $\ 23.15<m_a<24.0\ \mathrm{\mu eV}\ $ mass range. Phase II of the experiment will include upgrades to the cryogenics system and a new squeezed-state receiver. Finally, we discuss work on multi-rod cavities and photonic band gap resonators for higher frequency operation.

\end{abstract}

\section{Introduction}

Axions are a promising light dark matter candidate that also solve the Strong CP problem. They arise as the pseudogoldstone bosons that appear when a new global $U(1)_{PQ}$ symmetry is broken \cite{PecceiQuinnPRL,PecceiQuinnPRD,Weinberg,Wilczek}. 

\begin{figure}
\centerline{\includegraphics[width=0.84\textwidth]{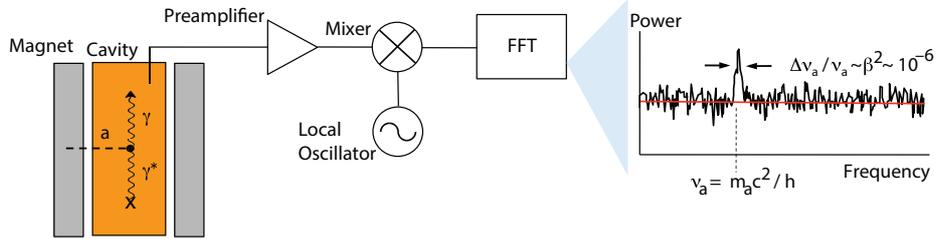}}
\caption{Schematic for haloscope with expected spectrum from axion event at $\nu_a$ with a bandwidth of $\Delta\nu_a$.}\label{haloscope}
\end{figure}

To-date, most experimental efforts to search for the axion have utilized the Primakoff effect in which the axion couples to a virtual photon provided by a strong magnetic field to produce a real detectable photon. In these searches, a microwave cavity is used to resonantly enhance the signal produced by the axion. This detector setup is called a haloscope. Such cavities are tunable so that the frequency of a resonant mode of interest matches the mass of the axion, i.e. $\nu=m_a c^2/h\ [1+\mathcal{O}(10^{-6})]$ \cite{Sikivie}. A schematic of this setup is shown in Fig. \ref{haloscope}.

The weak signal power of these cavities makes axion searches inherently difficult. The signal power is given by:
$$P=\left(\frac{g_\gamma^2\alpha^2\rho_a}{\pi^2\Lambda^4}\right)\left(\omega_cB_0^2VC_{nml}Q_L\frac{\beta}{1+\beta}\right)$$
where $g_\gamma$ is a model-dependent coupling constant, $\rho_a$ is the local dark matter density, set at $\rho\approx0.45\ \textrm{GeV/cm}^3$, and $\Lambda = 78\ $GeV. The second parenthesis contains experimental parameters: $\omega_c$ is the resonant frequency of the cavity, $B_0$ is the applied magnetic field, $V$ is the volume of the cavity, $C_{nml}$ is the mode form factor, and $Q_L=Q_0/(1+\beta)$, where $Q_0$ is the quality factor of the mode. Cavity designs focus on increasing the values for these latter parameters.

Since the power is very weak, searches focus on increasing the signal-to-noise ratio which is given by
$$SNR=\frac{P}{k_BT_S}\sqrt{\frac{t}{\Delta \nu_a}}$$
where $T_S$ is the system noise temperature, $\Delta\nu_a$ is the bandwidth of the signal line, and $t$ is the integration time. It is critical to be able to reach a low system noise temperature, which in turn is given by
$$k_BT_S=h\nu\left(\frac{1}{e^{h\nu/k_BT}-1}+\frac{1}{2}+N_A\right)$$
where $N_A$ is the noise from the amplifier, and $T$ is the physical temperature. As a result, when the temperature of the system is sufficiently low, the main source of noise is quantum in origin. Limiting this noise is a key component of improving axion searches.

\section{HAYSTAC}

The Haloscope At Yale Sensitive To Axion CDM (HAYSTAC) is a collaboration of Yale University, the University of Colorado, Boulder, and the University of California, Berkeley. The current cavity has a TM$_{010}$ mode frequency range of $3.4-5.8$ GHz which corresponds to axions with $14\ \mu\textrm{eV}<m_a<24\ \mu\textrm{eV}$.

\subsection{Phase I}
\begin{figure}
\centerline{\includegraphics[width=0.87\textwidth]{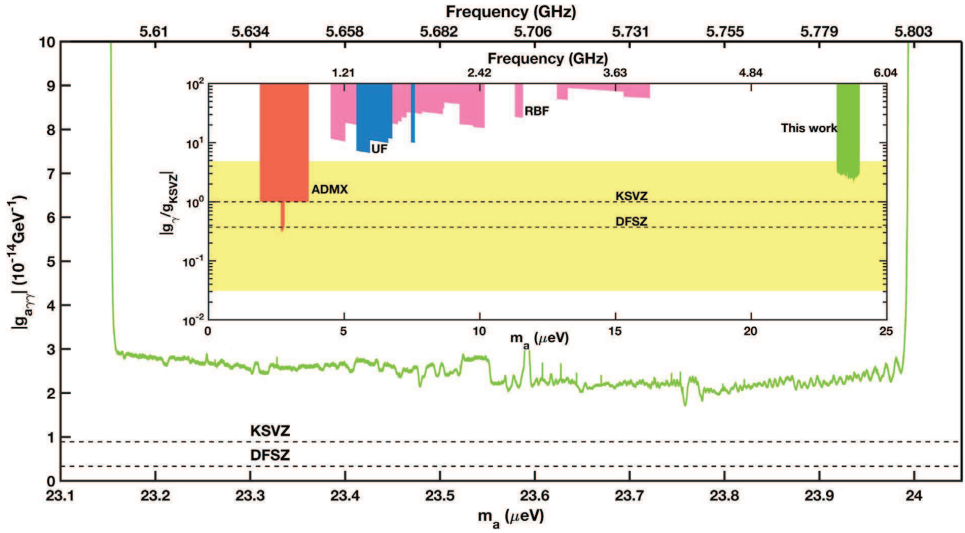}}
\caption{Exclusion limits set by Phase I of the HAYSTAC experiment in green \cite{PRD}.}\label{excl}
\end{figure}

The first phase of the HAYSTAC experiment, which was completed in 2018, excluded axions with $g_{a\gamma\gamma}\gtrsim2 \times 10^{-14}$ GeV in the $23.15<m_a<24.0\ \mu$eV range \cite{PRL,NIM,PRD}. This is the first experiment to exclude axions in the model band at masses greater than $10\ \mu$eV. These results are summarized in Fig. \ref{excl}.

This phase relied on a 2 L copper-plated stainless steel cylindrical cavity of 25.4 cm in height and 10.2 cm in diameter. A 5.1 cm diameter cylindrical copper tuning rod that pivots from the center of the cavity to the wall allows for the tuning of the TM$_{010}$ mode. This mode was chosen since its orientation along the z-axis maximized the form factor, which in turn is a measure of the overlap between the mode's electric field and the applied magnetic field. The magnetic field was provided by a 9 T magnet with a 14 cm diameter and 56 cm height bore.

The cavity was placed in a dilution refrigerator whose final plate was cooled to 127 mK. However, poor thermal contact between the rod body and the cold environment meant that the system was unable to cool below the standard quantum limit (SQL), which at 5 GHz corresponds to 240 mK. As a result, the first run was completed at $T\approx3\times T_{SQL}$ while an attempt to mitigate the problem in the second run lowered the temperature to $T\approx2\times T_{SQL}$. Further attempts to address this problem will be implemented in Phase II of the experiment.

A quantum limited Josephson Parametric Amplifier (JPA) was used for this phase. JPAs are nonlinear LC circuits whose inductance stems from an array of Superconducting Quantum Interference Devices (SQUIDs). The JPA provided a 20 dB gain and was tunable over 4.4-6.5 GHz. 

\subsection{Phase II}

Preparation for Phase II of the HAYSTAC experiment will primarily focus on two upgrades: enhanced thermal properties of the system, which include a new dilution refrigerator and improved thermal contact for the rod, and the introduction of a squeezed state receiver. 

A BlueFors Model LD 250 dilution refrigerator will be used for the next phase. By introducing this, the mechanical vibrations in the system will be reduced. To improve heat transfer between the tuning rod and the dilution refrigerator, two copper shafts are inserted into the two alumina alumina axles on either end of the rod. Tests at room temperature show that the copper shafts can be inserted sufficiently deep in the alumina tubes such that the thermal properties of the system are expected to significantly improve while the TM$_{010}$ only experiences a $\sim$$1-2$\% drop in the quality factor. 

\begin{figure}
\centerline{\includegraphics[width=\textwidth]{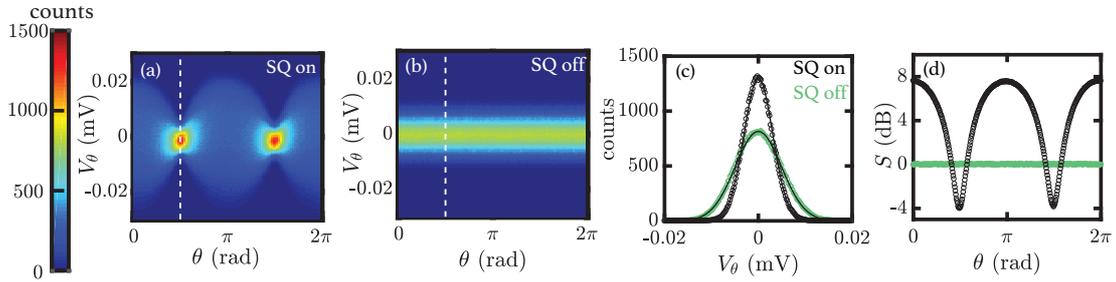}}
\caption{ Histogram for output voltage fluctuations as a function of the phase difference $\theta$ between the squeezer pump and the amplifier pump (a) with squeezing and (b) without squeezing. (c) Profiles of fluctuations along the dashed line of Figs. 3a and 3b ($\theta=\pi/2$) show that squeezing provides a less ambiguous signal. (d) Ratio of square of variances $S=\sigma_{\textrm{on}}^2/\sigma_{\textrm{off}}^2$ as a function of $\theta$ for the squeezed (black) and unsqueezed (green) states  \cite{PRApp}.}\label{ssr}
\end{figure}

Since the temperatures achieved by HAYSTAC are below the standard quantum limit, quantum noise becomes dominant. To address this issue, a squeezed state receiver will be used in Phase II. These devices rely on the uncertainty principle. At a given frequency the vacuum background can be written as $\mathbf{E_\omega}=\mathbf{E_0}(\hat{X}\mathrm{cos}(\omega t)+\hat{Y}\mathrm{sin}(\omega t))$ where $\hat{X}$ and $\hat{Y}$ are non-commuting observables. In the vacuum, the uncertainty can be modeled as being symmetric in phase space --- this leads to the histogram in Fig. \ref{ssr}b where the vacuum does not have a preferred orientation in phase space. The squeezed state receiver can then ``squeeze'' the state along one quadrature (at a phase difference $\theta$) thus reducing the uncertainty along that direction and leading to the histogram in Fig. \ref{ssr}a. This leads to an increase in the signal-to-noise ratio (SNR) of the system. Squeezing is only optimal when a wide range of frequencies is scanned, as in the case of haloscopes. If the exact frequency of the axion were known, a non-squeezed state would provide a higher SNR on resonance with the signal. However, by squeezing, the noise along the squeezed quadrature becomes weaker (Fig. \ref{ssr}c), therefore leading to shorter integration times to achieve the same SNR. This ultimately results in an increase in the scan rate of the system by a factor of 2.3.

\section{Future Cavity Designs}

Two new designs have been proposed for HAYSTAC's cavities: multi-rod cavities and photonic band gap structures. Multi-rod cavities act as a means of using the current cavity size and maintaining the same interaction volume while accessing higher frequencies. In particular, these cavities contain seven smaller rods: one stationary rod at the center of the cavity and six that surround the central rod. The six rods tune symmetrically outwards. This can be understood conceptually as if the seven rods act as an effective single larger rod whose radius is increasing while keeping its volume constant. Furthermore, simulations have shown that this design provides a higher form factor than that of a single larger tuning rod. This design will ultimately allow HAYSTAC to access higher frequencies in the future.

Photonic band gap (PBG) structures are also of particular interest since they can be designed to confine TM modes but not TE modes. They are similar to the current cavity but instead of a continuous wall, they contain a lattice of rods that acts as an effective wall. In the current cavity, TE modes plague regions of the frequency range since they mix with the TM modes due to the non-idealities of the cavity. When the TM$_{010}$ mode mixes with a TE mode, its form factor degrades, thereby rendering certain frequencies unusable. By using PBG structures, previously unusable frequencies will become accessible due to the lack of mode mixing. Forthcoming publications will discuss the PBG structures and the multi-rod cavities in detail.

\section{Acknowledgments}

This work was supported by NSF grants PHY-1362305 and PHY-1607417, Heising-Simons Foundation  grants 2014-181, 2014-182, and 2014-183, and U.S. DOE Contract DE-AC52-07NA27344. N. Rapidis is supported by the Haas Scholars Program.

\begin{footnotesize}

\begin{footnotesize}

\end{footnotesize}

\end{footnotesize}

\newpage

\end{document}